# Development of a higher order closed-form model for isotropic hyperelastic cylindrical body, including small vibration problem

E Hanukah

Faculty of Mechanical Engineering, Technion – Israel Institute of Technology, Haifa 32000, Israel

Email: eliezerh@tx.technion.ac.il

**Abstract**

We develop a new higher order closed-form model for the dynamics of a hyperelastic isotropic 3D cylindrical body. To this end, two material laws have been considered - Saint Venant-Kirchhoff and Neo-Hookean. We used systematic kinematic approximation, equivalent to Taylor multivariate expansion with respect to Cartesian coordinates, such that no axial or other symmetries are assumed enabling the model to capture 3D phenomena. Later, coordinates transformation imposed trigonometric terms to simple algebraic shape functions. Standard weak (Bubnov-Galerkin) form were adopted, while analytic integration resulted in a closed-form model - set of nonlinear coupled ordinary differential equations functions of geometrical, material and traction parameters. Internal forces which are built on Saint Venant-Kirchhoff material are exactly integrated in space. We demonstrated that neither numerical integration nor further approximations are necessary. Terms which are built on Neo-Hookean constitutive relation are systematically approximated which makes closed-form spatial integration possible. Importantly, we used consistency conditions with linear elasticity together with "exact" internal forces obtained by the use of Saint Venant-Kirchhoff model, to develop simple criterion that establishes the minimum order of approximation for Neo-Hookean. Our criterion has clear meaning of unphysical "zero" modes. Straightforward linearization of the system with respect to internal degrees of freedom leads to mass and stiffness matrices, such that small vibration problem becomes equivalent to algebraic eigenvalue problem. In particular, free vibration problem of 3D cylinder with equal height and diameter is considered. Natural frequencies based on first and second order of approximation are presented. In addition, we derive, for the first time, a simple explicit expression for the fundamental (lowest) frequency covering the entire range of practical engineering materials (positive Poisson's ratio) with high accuracy (0.03% error compared to FE results) based on a fourth order approximation.

## 1. Introduction

Many structural members take the form of cylindrical body with various parameters, e.g. rod-long cylinder, disc-short cylinder etc., therefore the development of dynamic models is of practical interest. Studying the vibrations of elastic three dimensional bodies is a well-known problem as well.

In the last several decades there has been an effort to provide 3-D elasticity solutions for the free vibration of prisms, parallelepipeds, cylinders etc., vast majority of these works have utilized various techniques to obtain *numerical results* (e.g. [1-18]).

Only a few studies applied structural theories that provide *closed-form* solutions for 3D elasticity problems (e.g. [19, 20]). These models take advantage of particular constitutive relations, justified by simplified assumptions on the structural behavior, that lead to a closed form equations of motion in terms of the degrees of freedom and also material and geometrical constants. The structural approaches which can be applied to 3-D problems can be roughly divided to two main classes: Pseudo Rigid Body and Cosserat Point, e.g. [21-28]. Cosserat Point method defines kinematic approximations in terms of directors, and also imposes restrictions on the strain energy function. In particular, the strain energy is separated into two parts; one uses average measures of deformation and the other is restricted to admit exact solutions, such as simple shear, simple torsion and pure bending, in an average sense. Cosserat point analytically satisfies Patch test, which ensures convergence when structure dimensions (size) tend to zero, see for example [29-32]. Pseudo-Rigid body is a Cosserat-like approach that enables closed form model for 3-D elastic solids of finite size. In addition to rigid body motion, it allows homogeneous deformation. Papadopoulos [33] has developed a second order theory of a pseudo-rigid body which has 30 degrees of freedom. It seems that the establishment of a pseudo-rigid body model on the basis of continuum mechanics is a delicate and unresolved issue (see [24, 26, 34]). The pseudo-rigid body method is used to simulate dynamics of multi-body systems which combine deformable solids with objects modeled as rigid bodies, e.g. [20, 35-40].

The present study follows basic guidelines in [41-43] together with convenient coordinate transformation, to formulate the governing equations of motion for a 3D cylinder. In particular, the 3-D elasticity problem is converted to a set of closed form non-linear ODEs. Similarly to finite element method - FEM (e.g.[44, 45]), especially its p-version (e.g.[46]), the formulation has two cornerstones; kinematic approximation and weak (Bubnov-Galerkin) form. Two essential differences distinguish this formulation from FEM; 1) We use kinematic approximation in terms of internal degrees of freedom, while in FEM, it's given in terms of nodes. Simple algebraic shape functions with respect to Cartesian coordinates are used, later transformation to cylindrical coordinates implies trigonometric terms. 2) Internal forces (and stiffness matrix) are integrated in closed-form manner, while in FEM numerical integration is used. Several studies show how to perform analytical integration for linear elasticity (e.g.[47-50]), however here nonlinear elasticity is discussed. In that sense, present study can be used by FEM too, which might result in significant time savings in computations, as it was established for linear theory (e.g.[47-50]). Saint Venant-Kirchhoff and Neo-Hookean constitutive laws have been considered.

Internal forces based on St. Venant-Kirchhoff are exactly integrated. Internal forces which are built on Neo-Hookean have been systematically approximated and then integrated. Consistency with linear elasticity is satisfied. Moreover, we suggest simple criterion that establishes the minimal order of approximation for Neo-Hookean, such that consistency with linear elasticity is satisfied. We specialize our model to examine the free-vibrating 3D cylinder with height equal to diameter. Although, comprehensive closed-form study is out of scope for present study, some meaningful results are derived. A highly accurate simple analytical expression for the fundamental (lowest) frequency based on a fourth order expansion is presented. It is emphasized that fundamental frequency is the most important for majority of engineering problems (e.g.[51] pp.303).

The outline of the paper is as follows. Section 2 presents the main theoretical considerations for deriving the non-linear dynamical equations of motion. In Section 2.1. Governing equations are obtained by adopting a weak formulation combined with analytical integration. Linearization of the governing equations with respect to the internal degrees of freedom is discussed in Section 3. Assuming small vibrations the mass and stiffness matrices are defined and an eigenvalue problem for the natural frequencies and modes is formulated. In Section 4, the solutions of free vibration problem are listed and discussed. The accuracy of closed-form expressions for the natural frequencies is examined by comparison to finite-elements simulations.

## 2. Theoretical considerations

Consider a three dimensional body occupying a finite volume in Euclidean space, which has a cylindrical shape in its reference configuration, made of isotropic and hyperelastic material. First, the basic equations of elasticity are recalled. (1) balance of linear momentum in initial configuration. Balance of angular momentum implies (2) where $\mathbf{P}$ is the first Piola-Kirchhoff stress tensor. The relation between Green-Lagrange strain tensor $\mathbf{E}$, and deformation gradient $\mathbf{F} = \text{Grad}(\mathbf{x})$ is given by (3), where gradient operator is defined by $\text{Grad}(\bullet) = \sum_{k=1}^{3} (\bullet)_{,k} \otimes \mathbf{G}^k$ and comma stand for partial differentiation with respect to coordinates, $\otimes$ stand for tensor/outer product. Constitutive stress-strain relations for St.Venant-Kirchhoff and Neo-Hookean materials are given by (4) and (5) respectively (e.g. [44] pp. 45):

$$\rho_0 \dot{\mathbf{v}} = \rho_0 \mathbf{b} + \text{Div}(\mathbf{P}) \tag{1}$$

$$\mathbf{P}\mathbf{F}^T = \mathbf{F}\mathbf{P}^T \tag{2}$$

$$\mathbf{E} = \frac{1}{2}\left(\mathbf{F}^T\mathbf{F} - \mathbf{I}\right) \tag{3}$$

$$\mathbf{P} = \mathbf{F}\left(\lambda(\mathbf{E}\bullet\mathbf{I})\mathbf{I} + 2\mu\mathbf{E}\right) \tag{4}$$

$$\mathbf{P} = \left(\frac{\lambda}{2}(J^2 - 1)\mathbf{I} + \mu\left(\mathbf{F}\mathbf{F}^T - \mathbf{I}\right)\right)\mathbf{F}^{-T} \tag{5}$$

Above, $\mathbf{X}$ and $\mathbf{x}$ represent the locations of a material point X in the initial and actual configurations, respectively, J denotes determinant of deformation gradient $J = \det(\mathbf{F})$, $\rho_0$ denotes initial mass density, $\lambda$ and $\mu$ are the Lame constants, $\mathbf{v} = \dot{\mathbf{x}}$ is the velocity of a material particle, a superposed dot denotes time differentiation, $\mathbf{b}$ stands for body force per unit of mass, divergence operator is given by $\mathrm{Div}(\bullet) = \sum_{k=1}^{3} (\bullet)_{,k} \mathbf{G}^k$, right transpose of arbitrary second order tensor $\mathbf{A}$ is denoted by $\mathbf{A}^T$, inverse of arbitrary invertible second order tensor $\mathbf{A}$ is denoted by $\mathbf{A}^{-1}$, dot ($\cdot$) stand for scalar product between two second order tensors (double contraction), $\mathbf{I}$ stands for second order identity tensor. Standard relations between material constants are recalled $\lambda = E\nu/((1+\nu)(1-2\nu)), \mu = E/2(1+\nu)$, where E and $\nu$ are the Young's modulus and Poisson's ratio. Restriction (2) is identically satisfied by material laws (4) and (5).

In present study two hyperelastic material laws are considered. St.Venant-Kirchhoff constitutive model is widely used in engineering applications especially for structural members like beams, rods or shells. It is referred to as generalization of Hook's law. Generally, it is applicable to deformations with large displacements and finite rotations but small strains. St.Venant-Kirchhoff's law do not satisfy the next conditions: in the limit case of the compression of a body to volume "0" the Cauchy stress $\boldsymbol{\sigma}$ approaches zero instead of $\lim_{J \to 0} \boldsymbol{\sigma} \to -\infty$. Therefore, it is not applicable for simulations of hyperelastic solids within finite deformation field [44] pp. 45. For example, rods, pipes, discs and cylindrical shell are special cases of the presented formulation. Neo-Hookean is one of the most widely used hyperelastic models. We decided to use both materials to demonstrate two different ways of implementing constitutive relation into the formulation. St.Venant-Kirchhoff due to its simplicity, it allows exact integration in the expressions of internal forces while Neo-Hookean will be systematically approximated to lead to closed-form expressions for internal forces and stiffness matrix. Since both material model consistent with linear elasticity for small strains about initial configuration, expressions built on St.Venant-Kirchhoff model will be used as a criterion to the to the expressions which are built on Neo-Hookean approximation. The treatment suggested here can be applied to any other hyperelastic constitutive relation. In this sense two of these models are good representatives of isotropic compressible hyperelasticity.

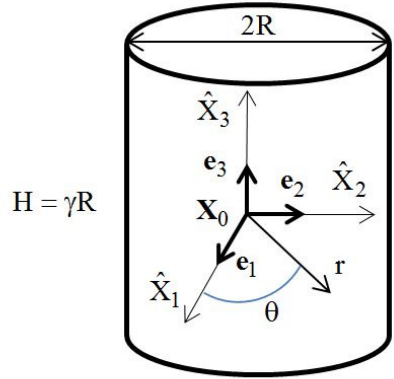

Figure 1: Schematic illustration showing the 3D finite cylinder, Cartesian orthonormal base vectors $\{\mathbf{e}_1, \mathbf{e}_2, \mathbf{e}_3\}$, their material coordinates $\{\hat{X}_1, \hat{X}_2, \hat{X}_3\}$, location of the centroid $\mathbf{X}_0$, cylindrical material coordinates $\{r, \theta, \hat{X}_3\}$, and geometrical parameters $R, \gamma, H$.

## 2.1 Kinematic approximation.

The initial geometry is defined by the means of centroid $\mathbf{X}_0$, three Cartesian base vectors $\{\mathbf{e}_1, \mathbf{e}_2, \mathbf{e}_3\}$; corresponding *material* coordinates $\{\hat{X}_1, \hat{X}_2, \hat{X}_3\}$ see Fig. 1. Material point $X \in \Omega_0$ occupying position $\mathbf{X}$ is exactly represented by

$$\mathbf{X} = \mathbf{X}_0 + \hat{X}_1 \mathbf{e}_1 + \hat{X}_2 \mathbf{e}_2 + \hat{X}_3 \mathbf{e}_3 \tag{6}$$

Using Lagrangian description, the actual configuration $\mathbf{x}$ of a material point $X$ is given by the motion $\mathbf{x} = \chi(\mathbf{X}(\hat{X}_1, \hat{X}_2, \hat{X}_3), t)$. The mapping function $\chi$ is unknown, and has to be determined. Finding the exact mapping is usually not possible therefore some approximations have to be considered. Here, we systematically approximate the motion function $\chi$ by means of Taylor's multivariable expansion about the centroid of the body $\mathbf{X}_0 = \chi(\mathbf{X}(0,0,0), 0)$, as follows

$$\mathbf{x}^h = \chi(\mathbf{X}, t)\big|_{\mathbf{X}=\mathbf{X}_0} + \sum_{k=1}^{3} \frac{\partial \chi(\mathbf{X}, t)}{\partial \hat{X}_k}\bigg|_{\mathbf{X}=\mathbf{X}_0} \hat{X}_k + \sum_{k,m=1}^{3} \frac{1}{2} \frac{\partial^2 \chi(\mathbf{X}, t)}{\partial \hat{X}_k \partial \hat{X}_m}\bigg|_{\mathbf{X}=\mathbf{X}_0} \hat{X}_k \hat{X}_m +$$
$$+ \sum_{k,m,n=1}^{3} \frac{1}{6} \frac{\partial^3 \chi(\mathbf{X}, t)}{\partial \hat{X}_k \partial \hat{X}_m \partial \hat{X}_n}\bigg|_{\mathbf{X}=\mathbf{X}_0} \hat{X}_k \hat{X}_m \hat{X}_n + \ldots \tag{7}$$

Here and throughout the text, upper symbol $()^h$ denotes approximated function so that $\mathbf{x}^h$ stands for approximation of $\mathbf{x}$. Partial derivatives of $\chi$ are evaluated at the centroid $\hat{X}_k = 0, (k=1,2,3)$, as a result they are merely functions of time and unknown. The order of approximation specifies the number of terms in the approximation (7) as well as the monomial terms $\hat{X}_k \hat{X}_m \hat{X}_n \ (k,m,n=0,..,3)$ which multiply the unknown (spatially constant) partial derivatives.

This forms a separation of variables where all the spatial dependence is in the monomials, and all the time dependence is in the derivatives. Next, inspired by (7), it is convenient to define the shape functions $N_i$ $(i=0,..,n^{shf})$

$$
\begin{aligned}
&N_0 = 1 \quad &&N_1 = \hat{X}_1 \quad &&N_2 = \hat{X}_2 \quad &&N_3 = \hat{X}_3 \quad &&N_4 = \hat{X}_1\hat{X}_2 \\
&N_5 = \hat{X}_1\hat{X}_3 \quad &&N_6 = \hat{X}_2\hat{X}_3 \quad &&N_7 = \hat{X}_1\hat{X}_1 \quad &&N_8 = \hat{X}_2\hat{X}_2 \quad &&N_9 = \hat{X}_3\hat{X}_3 \\
&N_{10} = \hat{X}_1\hat{X}_2\hat{X}_3 \quad &&N_{11} = \hat{X}_1\hat{X}_1\hat{X}_2 \quad &&N_{12} = \hat{X}_1\hat{X}_1\hat{X}_3 \quad &&N_{13} = \hat{X}_2\hat{X}_2\hat{X}_1 \quad &&N_{14} = \hat{X}_2\hat{X}_2\hat{X}_3 \\
&N_{15} = \hat{X}_3\hat{X}_3\hat{X}_1 \quad &&N_{16} = \hat{X}_3\hat{X}_3\hat{X}_2 \quad &&N_{17} = \hat{X}_1\hat{X}_1\hat{X}_1 \quad &&N_{18} = \hat{X}_2\hat{X}_2\hat{X}_2 \quad &&N_{19} = \hat{X}_3\hat{X}_3\hat{X}_3 \; .....
\end{aligned}
\qquad(8)
$$

Using the above definitions, together with the notion that all partial derivatives in (7) can be conveniently replaced by unknown time dependent vector terms $\mathbf{x}_j(t), (j=0,..,n^{shf})$, the kinematic approximation (7) is rewritten as

$$
\mathbf{x}^h = \sum_{j=0}^{n^{shf}} N_j \mathbf{x}_j(t)
$$

$$
1^{st} \text{ order} \Rightarrow n^{shf} = 3 \quad, \quad 2^{nd} \text{ order} \Rightarrow n^{shf} = 9 \qquad(9)
$$

$$
3^{rd} \text{ order} \Rightarrow n^{shf} = 19 \quad, \quad 4^{rd} \text{ order} \Rightarrow n^{shf} = 34 \; .....
$$

Unknown vector terms $\mathbf{x}_j(t), (j=0,..,n^{shf})$ have to be determined in order to estimate the deformation field. Each vector $\mathbf{x}_j(t)$ includes three components. Therefore the notion of internal degrees of freedom (IDF) arises naturally, and using (9) the total number of IDFs - $n^{dof}$, can be expressed by simple relation

$$
n^{dof} = 3(n^{shf} + 1). \qquad(10)
$$

Using definition (8), the exact representation (6) may be rewritten as

$$
\mathbf{X} = \sum_{j=0}^{n^{shf}} N_j \mathbf{X}_j \; , \; \mathbf{X}_j = \mathbf{e}_j \; (j=1,2,3) \; , \; \mathbf{X}_j\big|_{j>3} = 0 \qquad(11)
$$

To determine the initial values of $\mathbf{x}_j(t)$, it is recalled that approximation (9) has to admit (11) for the initial configuration, so the initial values of unknown functions $\mathbf{x}_j(t)$ are

$$
\mathbf{x}^h\big|_{\substack{t\to 0 \\ \chi\to \mathbf{1}}} = \mathbf{X} \;\Rightarrow\; \mathbf{x}_j\big|_{\substack{t\to 0 \\ \chi\to \mathbf{1}}} = \mathbf{X}_j \; , \; (j=0,..,n^{shf}) \qquad(12)
$$

where **1** stands for identity transformation. Next, displacement field will be approximated in the same manner, to this end it is convenient to define

$$
\mathbf{u}_j(t) = \mathbf{x}_j(t) - \mathbf{X}_j \; , \; \mathbf{u}_j\big|_{\substack{t\to 0 \\ \chi\to \mathbf{1}}} = \mathbf{0} \; , \; (j=0,..,n^{shf}). \qquad(13)
$$

Thus, the approximated displacement field becomes

$$\mathbf{u}^h = \mathbf{x}^h - \mathbf{X} = \sum_{j=0}^{n^{shf}} N_j \mathbf{u}_j(t) \tag{14}$$

Since (13) implies that all $\mathbf{u}_j(t)$, $(j=0,..,n^{shf})$ are zero in the initial configuration, it is natural to define the internal degrees of freedom as their components, i.e.

$$b_{3j+k}(t) = \mathbf{u}_j(t) \cdot \mathbf{e}_k \ , \ \mathbf{u}_j(t) = \sum_{k=1}^{3} b_{3j+k}(t) \mathbf{e}_k$$

$$b_m(t)\big|_{\substack{t \to 0 \\ \chi \to 1}} = 0 \ , \ (j=0,..,n^{shf}, m=1,..,n^{dof}) \tag{15}$$

Standard definitions of covariant and contravariant base vectors in initial configuration and some relations between them are recalled

$$\mathbf{G}_k = \partial \mathbf{X}/\partial \hat{X}_k = \mathbf{X}_{,k} = \mathbf{e}_k \ , \ (k=1,2,3) \ , \ G^{1/2} = \mathbf{G}_1 \times \mathbf{G}_2 \cdot \mathbf{G}_3 = 1 > 0 \tag{16}$$

$$\mathbf{G}^1 = \frac{\mathbf{G}_2 \times \mathbf{G}_3}{G^{1/2}} \ , \ \mathbf{G}^2 = \frac{\mathbf{G}_3 \times \mathbf{G}_1}{G^{1/2}} \ , \ \mathbf{G}^3 = \frac{\mathbf{G}_1 \times \mathbf{G}_2}{G^{1/2}} \tag{17}$$

Where $\delta_m^n$ is the Kronecker's delta, $(\cdot)$ stands for scalar product and $(\times)$ for vector product, comma $(\cdot)_{,i}$ $(i=1,2,3)$ denotes partial differentiation with respect to $\hat{X}_k$. Next we proceed to the following standard definitions of covariant and contravariant bases and their relations

$$\mathbf{g}_k = \partial \mathbf{x}^h/\partial \hat{X}_k = \mathbf{x}^h_{,k} \ , \ g^{1/2} = \mathbf{g}_1 \times \mathbf{g}_2 \cdot \mathbf{g}_3 > 0$$

$$\mathbf{g}^1 = \frac{\mathbf{g}_2 \times \mathbf{g}_3}{g^{1/2}} \ , \ \mathbf{g}^2 = \frac{\mathbf{g}_3 \times \mathbf{g}_1}{g^{1/2}} \ , \ \mathbf{g}^3 = \frac{\mathbf{g}_1 \times \mathbf{g}_2}{g^{1/2}} \ , \ \mathbf{g}_k \cdot \mathbf{g}^m = \delta_k^m \ , \ (k,m=1,2,3) \tag{18}$$

where comma denotes partial differentiation. Also, using the definitions of the deformation gradient, $\mathbf{F} = \dfrac{\partial \mathbf{x}}{\partial \mathbf{X}}$, together with (18),(16),(17) and (3), the approximated forms of the kinematic tensors are

$$\mathbf{F}^h = \sum_{k=1}^{3} \mathbf{g}_k \otimes \mathbf{G}^k \quad , \ (\mathbf{F}^h)^T = \sum_{k=1}^{3} \mathbf{G}^k \otimes \mathbf{g}_k \ ,$$

$$(\mathbf{F}^h)^{-1} = \sum_{k=1}^{3} \mathbf{G}_k \otimes \mathbf{g}^k \ , \ (\mathbf{F}^h)^{-T} = \sum_{k=1}^{3} \mathbf{g}^k \otimes \mathbf{G}_k \tag{19}$$

$$\tilde{\mathbf{F}}^{-T} = J^h (\mathbf{F}^h)^{-T} = \frac{1}{G^{1/2}} \sum_{k=1}^{3} \tilde{\mathbf{g}}^k \otimes \mathbf{G}_k \ , \ \tilde{\mathbf{g}}^m = g^{1/2} \mathbf{g}^m \ (m=1,2,3) \tag{20}$$

$$\mathbf{E}^h = \frac{1}{2}\left( \left(\mathbf{F}^h\right)^T \mathbf{F}^h - \mathbf{I} \right) \tag{21}$$

where $\otimes$ stands for tensor/outer product. Using the above, approximated constitutive laws (4) and (5) are given by

$$\mathbf{P}^{SVK} = \mathbf{P}^h = \mathbf{F}^h \left( \lambda \left( \mathbf{E}^h \bullet \mathbf{I} \right) \mathbf{I} + 2\mu \mathbf{E}^h \right) \qquad (22)$$

$$\mathbf{P}^{NH} = \mathbf{P}^h = \mu \mathbf{F}^h + \frac{\lambda}{2} J^h \tilde{\mathbf{F}}^{-T} - (\frac{\lambda}{2} + \mu) \frac{1}{J^h} \tilde{\mathbf{F}}^{-T} \qquad (23)$$

where ($\bullet$) stands for dot product between two second order tensors (double contraction). With the help of the above balance of linear momentum (1) is approximated as

$$\mathbf{R}^h = \text{Div}(\mathbf{P}^h) + \rho_0 \mathbf{b} - \rho_0 \dot{\mathbf{v}}^h \qquad (24)$$

where $\dot{\mathbf{v}}^h = \ddot{\mathbf{x}}^h$. Note, generally speaking, $\mathbf{R}^h$ does not vanish identically since kinematic approximation is implied. It is a function of $n^{dof}$ internal degrees of freedom $b_p(t)$, $(p=1,..,n^{dof})$. Next, we adopt a weak formulation to derive $(n^{shf}+1)$ vector equations of motion and consequently $n^{dof}$ scalar ODEs.

## 2.2 Equations of motion.

A weak formulation is used to restrict the residual $\mathbf{R}^h$ in the domain, and to obtain the governing dynamical equations of motion of the system

$$\mathbf{R}_i(R, H, E, \nu, \rho_0; b_p(t), \ddot{b}_q(t)) = \int_{\Omega_0} \mathbf{R}^h N_i dV = \mathbf{0} \;,\; (i=0,..,n^{shf}), (p,q=1,..,n^{dof}) \qquad (25)$$

The above weak form defines $(n^{shf}+1)$ vector coupled ordinary differential equations. With the help of partial integration, divergence theorem and introduction of the traction boundary condition $\bar{\mathbf{t}} = \mathbf{PN}$ ($\mathbf{N}$ is normal to the boundary in initial configuration), (25) is given by (e.g. [44] pp.84)

$$\int_{\Omega_0} N_i \rho_0 \dot{\mathbf{v}}^h dV = \int_{\Omega_0} N_i \rho_0 \mathbf{b} dV + \int_{\partial \Omega_0} N_i \bar{\mathbf{t}} dA - \int_{\Omega_0} \mathbf{P}^h \text{Grad}(N_i) dV \;,\; (i=0,..,n^{shf}) \qquad (26)$$

where $\text{Grad}(N_i) = \sum_{k=1}^{3} N_{i,k} \mathbf{G}^k$. Next, definitions for mass coefficients - $\hat{M}_{ij}$, internal forces - $\mathbf{f}_i^{int}$, external and body forces - $\mathbf{f}_i^{ext}, \mathbf{f}_i^{body}$, are introduced

$$\mathbf{f}_i^{ext} = \int_{\partial \Omega_0} N_i \bar{\mathbf{t}} dA$$

$$\mathbf{f}_i^{body} = \int_{\Omega_0} N_i \rho_0 \mathbf{b} dV \;,\; (i=0,..,n^{shf}) \qquad (27)$$

$$\hat{M}_{ij} = \rho_0 \int_{\Omega_0} N_i N_j dV$$

$$\mathbf{f}_i^{int} = \int_{\Omega_0} \mathbf{P}^h \text{Grad}(N_i) dV \;,\; (i,j=0,..,n^{shf}) \qquad (28)$$

With the help of (28),(14) and (9), the (vector) equations of motion (26), are written as follows

$$\mathbf{R}_i = \sum_{j=0}^{n^{shf}} \hat{M}_{ij}\ddot{\mathbf{u}}_j + \mathbf{f}_i^{int} - \mathbf{f}_i^{ext} - \mathbf{f}_i^{body} = \mathbf{0} \ , \ \left(i = 0,..,n^{shf}\right) \tag{29}$$

Importantly, it is possible to write explicit closed-forms of equations $\mathbf{R}_i$ in terms of the initial geometry parameter $R$, and the material constants $E, \nu, \rho_0$ and scalar functions $\ddot{b}_p(t), b_q(t)$, $(p,q = 1,..,n^{dof})$.

## 2.3 Closed-form integration.

Throughout the present study we do not use explicit body force $\mathbf{b}$ or external traction $\bar{\mathbf{t}}$ therefore expresions (27) are left untouched. However, mass coefficients $\hat{M}_{ij}$ and internal forces $\mathbf{f}_i^{int}$ require more detailes. The idea of closed form integration for nonlinear constitutive equations in (28) is, as far as we know, unique to this study. We present a systematic approach using which closed-form integration is always possible for any hyperelastic constitutive equation. Consider the next integral

$$\bar{\mathbf{I}} = \int_{\Omega_0} \hat{\mathbf{I}}(\hat{X}_1,\hat{X}_2,\hat{X}_2,t)dV \ , \ dV = \underbrace{G^{1/2}}_{1}d\hat{X}_1 d\hat{X}_2 d\hat{X}_3 = d\square \tag{30}$$

An integrand is a function of material coordinates and time. If an integrand $\hat{\mathbf{I}}(\hat{X}_1,\hat{X}_2,\hat{X}_2,t)$ can be exactly represented as it's multivariable expansion about the centroid (7), then using definition of shape functions (8), and renaming the timedependent partial derivatives one can conveniently rewrite

$$\hat{\mathbf{I}}(\hat{X}_1,\hat{X}_2,\hat{X}_2,t) = \sum_{i=0}^{\bar{n}} N_i \hat{\mathbf{I}}_i(t) \ , \ (\bar{n} = 0,1,2...) \tag{31}$$

Therefore (30) is given by

$$\bar{\mathbf{I}} = \sum_{i=0}^{\bar{n}} \hat{\mathbf{I}}_i(t) \int_{\Omega_0} N_i(\hat{X}_1,\hat{X}_2,\hat{X}_3)d\square \tag{32}$$

Integration in cylindrical domain imply that coordinates $\{\hat{X}_1,\hat{X}_2,\hat{X}_3\}$ are changed to Cylindrical coordinates $\{r,\theta,\hat{X}_3\}$ see Fig. 1. by the next transformation

$$\hat{X}_1 = r\cos(\theta) \ , \ \hat{X}_2 = r\sin(\theta) \ , \ dV = rdrd\theta d\hat{X}_3 \tag{33}$$

Using the above transformation, integral (32) is given by

$$\bar{\mathbf{I}} = \sum_{i=0}^{\bar{n}} \hat{\mathbf{I}}_i(t) \int_{-\frac{H}{2}}^{+\frac{H}{2}} \int_0^{2\pi} \int_0^R N_i(r,\theta,\hat{X}_3)rdrd\theta d\hat{X}_3 \tag{34}$$

Expression (34) is easily analytically integrated, commercial software MAPLE$^{TM}$ were extensively used throughout this study.

Accordingly to (34), it is easy to see that mass coefficient $\hat{M}_{ij}$ given by (28), are simple to compute.

$$\hat{M}_{ij} = \rho_0 \int_{-\frac{H}{2}}^{+\frac{H}{2}} \int_0^{2\pi} \int_0^R N_i(r,\theta,\hat{X}_3) N_j(r,\theta,\hat{X}_3) r \, dr \, d\theta \, d\hat{X}_3 \, , \, (i,j=0,..,n^{shf}) \quad (35)$$

Internal forces $\mathbf{f}_i^{int}$ which are given by (28) divided to two cases, internal forces built on Saint Venant-Kirchhoff material (22), and internal forces that take into account Neo-Hookean material (23) such that

$$\mathbf{f}_i^{SVK} = \mathbf{f}_i^{int}\Big|_{\mathbf{P}^h = \mathbf{P}^{SVK}} = \int_{\Omega_0} \mathbf{P}^{SVK} \text{Grad}(N_i) dV$$

$$\mathbf{f}_i^{NH} = \mathbf{f}_i^{int}\Big|_{\mathbf{P}^h = \mathbf{P}^{NH}} = \int_{\Omega_0} \mathbf{P}^{NH} \text{Grad}(N_i) dV \, , \, (i,j=0,..,n^{shf}) \quad (36)$$

Importantly, $\mathbf{f}_i^{SVK}$ take the form (34), hence they are exactly integrated. To clarify this point, we denote the order of expansion (7) by $\alpha$, with the help of (18) it follows that $\mathbf{g}_k$ are $\alpha-1$ order with respect to coordinates, next using (16),(17),(19) it is clear that $\mathbf{F}^h$ is $\alpha-1$ order with respect to coordinates, next using (21) we obtain that $\mathbf{E}^h$ is $2(\alpha-1)$ order, hence using (22) we see that $\mathbf{P}^{SVK}$ is $3(\alpha-1)$ with respect to coordinates. Next we return to (36) and note that $\text{Grad}(N_i) = \sum_{k=1}^{3} N_{i,k} \mathbf{G}^k$ is $\alpha-1$ order of coordinates, therefore the integrand of $\mathbf{f}_i^{SVK}$ given in (36) is $4(\alpha-1)$ order with respect to coordinates, hence, takes the form (34). Which means that for chosen order of kinematic approximation (7) and as a result (9), denoted by $\alpha$, one can exactly represent $\mathbf{P}^{SVK}$ given by (22), in a separate form (31), which together with coordinates transformation (33) leads to exactly integrated form (34).

According to the above, $\mathbf{f}_i^{SVK}$ are exactly computed. Unfortunately, the same cannot be done for $\mathbf{f}_i^{NH}$, since $\mathbf{P}^{NH}$ given by (23) is rational function with respect to coordinates. To illustrate this point, we once again denote the order of kinematic approximation by $\alpha$, using (18) we get that $g^{1/2}$ is $3(\alpha-1)$ order ($J^h = g^{1/2}/G^{1/2}$ is also of order $3(\alpha-1)$), so $\mathbf{g}^k$ is a rational function with respect to coordinates but $\tilde{\mathbf{g}}^k = g^{1/2}\mathbf{g}^k$ turns out to be order $2(\alpha-1)$, next following the definition (20) we see that $\tilde{\mathbf{F}}^{-T}$ is also $2(\alpha-1)$, which bring us to approximated first Piola-Kirchhoff stress tensor for Neo-Hookean material - $\mathbf{P}^{NH}$ given by (23). The first term in (23) is $\mu \mathbf{F}^h$, which is $(\alpha-1)$ order of coordinates, the second term $\frac{\lambda}{2}J^h\tilde{\mathbf{F}}^{-T}$ is of order $5(\alpha-1)$, and the last term $(\frac{\lambda}{2}+\mu)\frac{1}{J^h}\tilde{\mathbf{F}}^{-T}$ is clearly a rational function of coordinates which means that the integrand cannot be exactly represented with its multivariable expansion about the centroid using finite order and straightforward exact integration (36) is, generally speaking, impossible. In present study, we approximate $\mathbf{P}^{NH}$ by its Taylor's multivariable expansion about the centroid $\mathbf{X}_0$. The function $\mathbf{P}^{NH}$ is analytic therefore its Taylor series converge, and convenient form (31) obtained, coordinates transformation (33) is applied and integration is performed. To this end, one has to choose the order of approximation, keeping in mind that as an order increases, the approximations becomes better together with growing computational effort, since higher order means more terms to compute. Therefore, it is important to establish the minimum order of expansion of $\mathbf{P}^{NH}$. The first terms of (23) has $(\alpha-1)$ order (the second $8(\alpha-1)$ and the third is rational) and this is the minimum of suggested order of approximation. Approximating $\mathbf{P}^{NH}$ to $(\alpha-1)$ order and using (36),(33),(34) internal forces are computed.

To establish the optimal order of expansion for $\mathbf{P}^{NH}$ in the sense of accuracy, robustness, computational effort etc., comprehensive numerical study need to be conducted. Nevertheless, hare we provide a way to ensure derivation of closed-form model, way which can be generalized to any hyperelastic constitutive relations, and we provide the order below which unphysical behavior is guaranteed. For linearized theory, both $\mathbf{f}_i^{SVK}$ and $\mathbf{f}_i^{NH}$ should be equal; $\lin(\mathbf{f}_i^{SVK}) = \lin(\mathbf{f}_i^{NH}), (i = 0,..,n^{shf})$. In the next section we verify that using lower than $(\alpha-1)$ order to approximate $\mathbf{P}^{NH}$ result in internal forces that do not satisfy this condition, unphysical "zero" modes for small vibration modes appear.

## 3. Linearization and formulation of the free vibration problem

In the previous section, vector equation of motion has been derived. Herein, derivation of a set of $n^{dof}$ scalar equations of motion is presented, linearization is carried out, mass and stiffness matrices are defined, and an eigenvalue problem is formulated for the free vibration analysis. To this end, we define

$$F_{3i+k}\left(R, H, E, \nu, \rho_0; \ddot{b}_p(t), b_q(t)\right) = \mathbf{R}_i \bullet \mathbf{e}_k = 0$$
$$\left(i = 0,..,n^{shf}\right), \left(k = 1, 2, 3\right), \left(p, q = 1,..,n^{dof}\right) \tag{37}$$

and

$$F_{3i+k}^{mass}\left(R,H,\rho_0;\ddot{b}_p(t)\right) = \sum_{j=0}^{n^{shf}} \hat{M}_{ij}\ddot{u}_j \cdot e_k$$

$$F_{3i+k}^{stiff}\left(R,H,E,\nu;b_p(t)\right) = f_i^{int} \cdot e_k \tag{38}$$

$$\left(i=0,..,n^{shf}\right),\left(k=1,2,3\right),\left(p=1,..,n^{dof}\right)$$

Free-vibration problem is investigated, therefore external and body forces are neglected. These enable us to compactly rewrite (37) as

$$F_p = F_p^{mass} + F_p^{stiff} = 0 \ , \ \left(p=1,..,n^{dof}\right). \tag{39}$$

Next, we introduce the following notations

$$[\ddot{b}] = \begin{bmatrix} \ddot{b}^1 \\ \cdot \\ \cdot \\ \ddot{b}^{n^{dof}} \end{bmatrix}, \ [b] = \begin{bmatrix} b^1 \\ \cdot \\ \cdot \\ b^{n^{dof}} \end{bmatrix}, \ [F] = [F^{mass}] + [F^{stiff}] = \begin{bmatrix} F_1^{mass} \\ \cdot \\ \cdot \\ F_{n^{dof}}^{mass} \end{bmatrix} + \begin{bmatrix} F_1^{stiff} \\ \cdot \\ \cdot \\ F_{n^{dof}}^{stiff} \end{bmatrix} = [0] \tag{40}$$

Linearization of the system $[F]=[0]$ is carried out about the initial configuration, namely $[b]=[0]$, such that $[F] \approx [F]_{[b]=[0]} + \dfrac{\partial[F]}{\partial[b]}\bigg|_{[b]=[0]} [b]$. Using definitions(38), notation(40), the mass and stiffness matrices of the system of ODE (39) can written as:

$$[K] = \begin{bmatrix} K_{11} & . & K_{1n^{dof}} \\ . & . & . \\ K_{n^{dof}1} & . & K_{n^{dof}n^{dof}} \end{bmatrix} \ , \ K_{ij} = \dfrac{\partial F_i^{stiff}}{\partial b_j}\bigg|_{[b]=[0]}$$

$$[M] = \begin{bmatrix} M_{11} & . & M_{1n^{dof}} \\ . & . & . \\ M_{n^{dof}1} & . & M_{n^{dof}n^{dof}} \end{bmatrix} \ , \ M_{ij} = \dfrac{\partial F_i^{mass}}{\partial \ddot{b}_j} \tag{41}$$

Both matrices are symmetric. In addition, the mass matrix is positive definite and the stiffness matrix is positive semi-definite due to rigid body motion modes. Using the above definition, the linearization of the system (39) about the initial configuration $[b]=[0]$ is given by

$$[F] \approx [M][\ddot{b}] + [K][b] = [0] \tag{42}$$

Next it is assumed that $[b] = [\tilde{b}]\sin(\omega t)$ were $\omega$ is the vibration natural frequency and $[\tilde{b}]$ is the mode of the vibration (constant algebraic vector). The second time derivative becomes $[\ddot{b}] = -\omega^2[\tilde{b}]\sin(\omega t)$, and after substituting to (42) the well-known form for small vibration problem in many degree of freedom (MDOF) system is

$$\left(-\omega^2[M] + [K]\right)[\tilde{b}]\sin(\omega t) = [0] \tag{43}$$

Non-trivial solutions for $[\tilde{b}]$ exists if and only if $\left|-\omega^2[M]+[K]\right|=0$ (determinant vanishes). Solution of this problem leads to $n^{dof}$ natural modes and frequencies. The lowest six natural frequencies are zero, and represent rigid body translation and rotation.

Importantly, due to consistency with linear elasticity, stiffness matrix built $\mathbf{f}_i^{SVK}$ and stiffness matrix built on $\mathbf{f}_i^{NH}$ have to be equal. If order of approximation for $\mathbf{P}^{NH}$ is lower than $(\alpha-1)$ then consistency is not maintained, and non-physical zero modes appear in free vibration problem.

## 4. Natural frequencies - 3D cylinder: $\gamma = 2$

The main objective of the below chapter is to explicitly show that the above formulation can be used to produce new, simple, closed-form results with significant engineering importance. Comprehensive closed-form study of small-vibration problem is out of scope of current study. Herein, in order to demonstrate the ability to capture truly 3D modes, parameter $\gamma$ was set to 2, such that the width and height of the discussed structure will be the same.

All the expressions for resonant frequencies for all orders which has been addressed (1st-4st) throughout the research take the next form

$$\omega_k^2 = \frac{E}{\rho_0 R^2} \bar{\omega}_k^2(\nu) \ , \ (k=1,2,3...) \tag{44}$$

where $\bar{\omega}_k$ is a non-dimensional frequency. This result provides an essential practical insight regarding the role of material and geometrical properties of a free vibrating cylinder. Finally, we note that although some of the below expressions involve complex terms, *all frequencies are real*. This is a direct consequence of the fact that the mass and stiffness matrices are positive definite and semi-positive definite, respectively, in addition to both being real and symmetric (e.g.[51] pp. 293).

**First order approximation.** We begin by calculating the natural frequencies associated with the first order approximation, using (18),(19) leads to deformation gradient is homogeneous $\mathbf{F}^h = \mathbf{F}^h(t)$. To this end, the stiffness and mass matrices (41) are formulated and the eigenvalue problem (43) is solved. First order i.e. $n^{dof}=12$, and therefore the number of natural frequencies is 12. Also, as stated earlier, the first 6 natural frequencies match rigid body motion modes (three rigid body translation and three rotation modes) are zero. The remaining 6 non-trivial natural frequencies are

$$\bar{\omega}_1^2 = \frac{-7+3\nu+\sqrt{1+6\nu+105\nu^2}}{-2+2\nu+4\nu^2}$$

$$\bar{\omega}_2^2 = \bar{\omega}_3^2 = \frac{7}{2(1+\nu)}$$

$$\bar{\omega}_4^2 = \bar{\omega}_5^2 = \frac{4}{1+\nu} \tag{45}$$

$$\bar{\omega}_6^2 = \frac{-7+3\nu-\sqrt{1+6\nu+105\nu^2}}{-2+2\nu+4\nu^2}$$

All the above eigenvalues (squared normalized natural frequencies) are associated with homogeneous deformation modes (eigenvectors). In (45) there are two frequencies which are repeated twice. Linear combinations of the modes that belong to the same eigenvalue are eigenvectors as well. Next, we excel this solution by providing better accuracy and richer spectral analysis by means of higher order approximations.

**Second order approximation.** The second order approximation involves $n^{dof} = 30$, namely the system has $30-6=24$ non-trivial frequencies, four times more than the previous solution (which provided 6 nontrivial expressions (four of them are distinct). Formulating (41) and solving the corresponding eigenvalue problem (43), we find that the first order solution - (45), is included in the second order solution, the additional natural frequencies are given by

$$\bar{\omega}_7^2 = \frac{3}{2(1+\nu)} \ , \ \bar{\omega}_8^2 = \bar{\omega}_9^2 = \frac{23-\sqrt{145}}{4(1+\nu)} \ , \ \bar{\omega}_{10}^2 = \bar{\omega}_{11}^2 = \frac{23+\sqrt{145}}{4(1+\nu)} \ , \ \bar{\omega}_{12}^2 = \bar{\omega}_{13}^2 = \frac{12}{1+\nu} \tag{46}$$

$$\bar{\omega}_{14}^2 = \frac{r_3^{1/3}}{6(1+v)^2(-1+2v)^2} + \frac{-998+14544v^4+240v^3-8792v^2+4514v}{12(-1+v+2v^2)r_3^{1/3}} + \frac{84v-65}{-6+6v+12v^2}$$

$$\bar{\omega}_{15}^2 = \frac{(-92v^2+336v^3+130-298v)(-(1+v)^3 r_4(-1+2v)^3)^{1/3}}{12r_5} +$$

$$+ \frac{(i\sqrt{3}-1)8^{2/3}((-1/2+v)^3(1+v)^3 r_2)^{2/3}}{12r_5} -$$

$$- \frac{-14544(1+i\sqrt{3})\left(\frac{499}{3636}-\frac{293}{606}v+v^2\right)(-1/2+v)^2(1+v)^2}{12r_5}$$

$$\bar{\omega}_{16}^2 = \frac{(-92v^2+336v^3+130-298v)(-(1+v)^3 r_4(-1+2v)^3)^{1/3}}{12r_5} -$$

$$- \frac{8^{2/3}(1+i\sqrt{3})((-1/2+v)^3(1+v)^3 r_2)^{2/3}}{12r_5} +$$

$$+ \frac{14544\left(\frac{499}{3636}-\frac{293}{606}v+v^2\right)(-1/2+v)^2(1+v)^2(i\sqrt{3}-1)}{12r_5}$$

$r_5 = (-(1+v)^3 r_4(-1+2v)^3)^{1/3}(1+v)^2(-1+2v)^2$

$r_4 = -18\sqrt{r_1}v + 9\sqrt{r_1} + 149256v^3 - 111942v^2 + 64791v - 10900$

$r_3 = r_2(1+v)^2(-1+2v)^2(-1+v+2v^2)$

$r_2 = -149256v^3 + 111942v^2 - 64791v + 18\sqrt{r_1}v + 10900 - 9\sqrt{r_1}$

$r_1 = -79606080v^4 + 32458320v^3 - 14402667v^2 - 1493562v - 67179$

$$\bar{\omega}_{17} = \bar{\omega}_{18} = \frac{92v\hat{r}_3^{1/6}\hat{r}_4^{1/4} - 73\hat{r}_3^{1/6}\hat{r}_4^{1/4} + \hat{r}_4^{3/4} + \hat{r}_5^{1/2}}{8\hat{r}_3^{1/6}\hat{r}_4^{1/4}(1+v)(-1+2v)}$$

$$\bar{\omega}_{19} = \bar{\omega}_{20} = \frac{92v\hat{r}_3^{1/6}\hat{r}_4^{1/4} - 73\hat{r}_3^{1/6}\hat{r}_4^{1/4} + \hat{r}_4^{3/4} - \hat{r}_5^{1/2}}{8\hat{r}_3^{1/6}\hat{r}_4^{1/4}(1+v)(-1+2v)}$$

$$\bar{\omega}_{21} = \bar{\omega}_{22} = \frac{92v\hat{r}_3^{1/6}\hat{r}_4^{1/4} - 73\hat{r}_3^{1/6}\hat{r}_4^{1/4} - \hat{r}_4^{3/4} + \hat{r}_6^{1/2}}{8\hat{r}_3^{1/6}\hat{r}_4^{1/4}(1+v)(-1+2v)}$$

$$\bar{\omega}_{23} = \bar{\omega}_{24} = \frac{92v\hat{r}_3^{1/6}\hat{r}_4^{1/4} - 73\hat{r}_3^{1/6}\hat{r}_4^{1/4} - \hat{r}_4^{3/4} - \hat{r}_6^{1/2}}{8\hat{r}_3^{1/6}\hat{r}_4^{1/4}(1+v)(-1+2v)}$$

$$\hat{r}_6 = 28160r2v^3 - 381712\sqrt{r4}v^3 + 103296r2v^2 + 6240\sqrt[3]{r3}\sqrt{r4}v^2 + 489640\sqrt{r4}v^2 - 16r3^{2/3}\sqrt{r4}v -$$
$$-4688\sqrt[3]{r3}\sqrt{r4}v - 61560r2v - 273056\sqrt{r4}v + 2242\sqrt[3]{r3}\sqrt{r4} + 8r3^{2/3}\sqrt{r4} + 61832\sqrt{r4} + 40802r2$$
$$\hat{r}_5 = -28160r2v^3 - 381712\sqrt{r4}v^3 + 6240\sqrt[3]{r3}\sqrt{r4}v^2 + 489640\sqrt{r4}v^2 - 103296r2v^2 - 16r3^{2/3}\sqrt{r4}v -$$
$$-4688\sqrt[3]{r3}\sqrt{r4}v + 61560r2v - 273056\sqrt{r4}v + 2242\sqrt[3]{r3}\sqrt{r4} - 40802r2 + 8r3^{2/3}\sqrt{r4} + 61832\sqrt{r4}$$
$$\hat{r}_4 = 381712v^3 + 3120\sqrt[3]{r3}v^2 - 489640v^2 + 16r3^{2/3}v - 2344\sqrt[3]{r3}v + 273056v - 61832 - 8r3^{2/3} + 1121\sqrt[3]{r3}$$
$$\hat{r}_3 = 3294935v^3 - 4206957v^2 + 2485113v - 655307 + 6r1$$
$$\hat{r}_2 = \sqrt{3294935v^3 - 4206957v^2 + 2485113v - 655307 + 6r1}$$
$$\hat{r}_1 = (-75604687488v^6 + 115610635008v^5 - 113335764150v^4 + 53997989244v^3 - 18658769196v^2 +$$
$$+ 2488528620v - 896761590)^{1/2}$$

The lowest resonant frequency is called fundamental frequency, for most engineering cases it is an important one (e.g.[51] pp.303). Thus, the solution provided by the second order approximation is superior to the first order solution since it leads to better representation of fundamental frequency. Moreover, fifteen different formulas are observed instead of four, which means that second order model significantly enriches response spectrum of the structure.

Fig. 2 shows the dependence of the squares of natural frequencies on poison's ratio - $v$, for the second order solution. The first order frequencies are illustrated by the gold lines. It is evident that the first order approximation is not able to capture the lowest frequencies. Moreover, the third order solution demonstrates that different modes may be of greater engineering importance depending on Poisson's ratio.

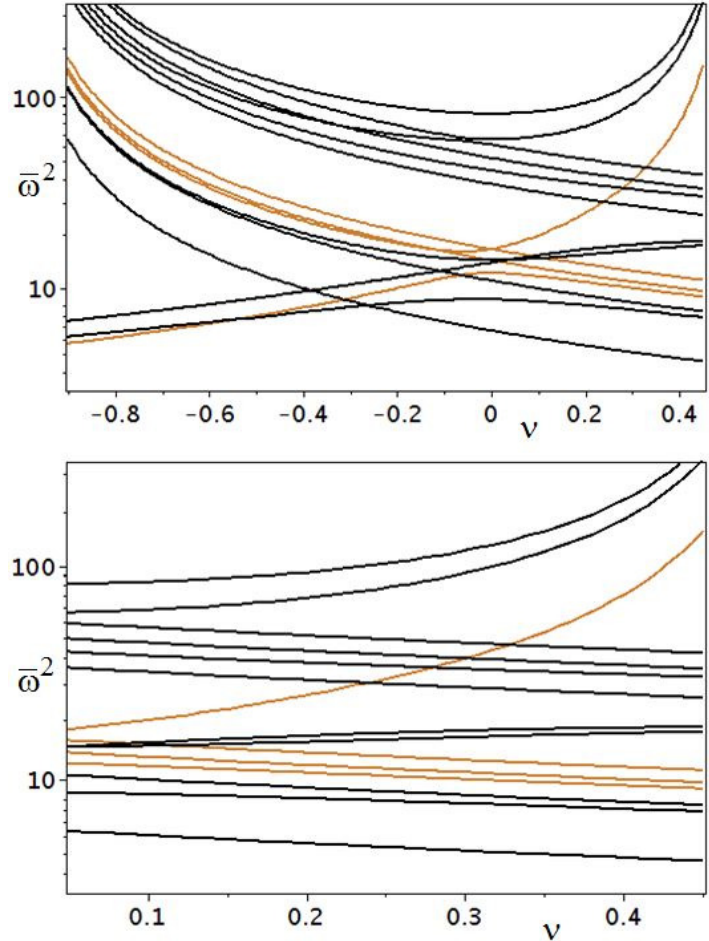

Figure 2: Second order solution showing square non-trivial natural frequencies of a free vibrating sphere as a function of Poisson's ratio. Also, frequencies which correspond to the first order approximation, are illustrated in gold. The lower plot is identical to the upper, but focuses on the engineering range of $\nu > 0$.

Next, we examine the accuracy of the approximated solutions. To this end, we compare between the 24 lowest frequencies (excluding the six trivial modes associated with rigid body motion) calculated by first, second, third and fourth order approximations to the results of a finite elements method (FEM). The finite elements analysis was performed with the commercial software ABAQUS$^{TM}$ 6.10, using standard 20 node full integration brick element *C3D20*, and for the following typical parameters $E = 210[GPa]$, $\rho_0 = 7800[Kg/m^3]$, $\nu = 0.3$, $R = 1/2[m]$. However, Table 1 consists of $\bar{\omega}$ values for convenience.

Converges of the finite element analysis was validated by increasing the number of elements and examining the tenth lowest natural frequency. FEM results provided in Table 1 were calculated using 16350 elements, and are found to be fully converged.

| Mode # | FEM | O(1) | O(2) | O(3) | O(4) | Err % O(1) | Err % O(2) | Err % O(3) | Err % O(4) |
|---|---|---|---|---|---|---|---|---|---|
| 1 | 1.948 | 3.162 | 2.148 | 2.148 | 1.949 | 62.31 | 10.27 | 10.27 | 0.03 |
| 2 | 2.454 | 3.282 | 2.775 | 2.733 | 2.474 | 33.73 | 13.09 | 11.39 | 0.83 |
| 3 | 2.454 | 3.282 | 2.775 | 2.733 | 2.474 | 33.73 | 13.09 | 11.39 | 0.83 |
| 4 | 2.473 | 3.508 | 2.903 | 2.775 | 2.491 | 41.86 | 17.40 | 12.21 | 0.71 |
| 5 | 2.473 | 3.508 | 2.903 | 2.775 | 2.491 | 41.86 | 17.40 | 12.21 | 0.71 |
| 6 | 2.668 | 6.076 | 3.162 | 2.899 | 2.733 | 127.75 | 18.52 | 8.64 | 2.45 |
| 7 | 2.668 | --- | 3.282 | 2.903 | 2.733 | --- | 23.00 | 8.82 | 2.45 |
| 8 | 2.885 | --- | 3.282 | 2.903 | 2.899 | --- | 13.73 | 0.62 | 0.46 |
| 9 | 2.900 | --- | 3.508 | 2.918 | 2.918 | --- | 20.97 | 0.62 | 0.62 |
| 10 | 2.900 | --- | 3.508 | 2.918 | 2.918 | --- | 20.97 | 0.62 | 0.62 |
| 11 | 3.086 | --- | 3.967 | 3.435 | 3.435 | --- | 28.55 | 11.33 | 11.33 |
| 12 | 3.086 | --- | 3.967 | 3.435 | 3.435 | --- | 28.55 | 11.33 | 11.33 |
| 13 | 3.475 | --- | 4.109 | 3.967 | 3.495 | --- | 18.23 | 14.16 | 0.59 |
| 14 | 3.475 | --- | 5.192 | 3.967 | 3.495 | --- | 49.41 | 14.16 | 0.59 |
| 15 | 3.562 | --- | 5.192 | 4.109 | 3.618 | --- | 45.76 | 15.34 | 1.56 |
| 16 | 3.804 | --- | 5.739 | 4.173 | 4.149 | --- | 50.85 | 9.69 | 9.05 |
| 17 | 3.822 | --- | 5.739 | 4.173 | 4.149 | --- | 50.15 | 9.19 | 8.55 |
| 18 | 3.822 | --- | 6.076 | 4.406 | 4.173 | --- | 58.99 | 15.27 | 9.19 |
| 19 | 3.897 | --- | 6.076 | 4.804 | 4.173 | --- | 55.94 | 23.28 | 7.10 |
| 20 | 3.957 | --- | 6.076 | 5.096 | 4.406 | --- | 53.58 | 28.81 | 11.35 |
| 21 | 3.957 | --- | 6.605 | 5.096 | 4.451 | --- | 66.93 | 28.81 | 12.49 |
| 22 | 3.960 | --- | 9.622 | 5.138 | 4.451 | --- | 143.00 | 29.76 | 12.41 |
| 23 | 3.960 | --- | 10.874 | 5.192 | 4.525 | --- | 174.63 | 31.12 | 14.27 |
| 24 | 4.138 | --- | 10.874 | 5.192 | 4.525 | --- | 162.75 | 25.45 | 9.33 |

Table 1: Comparison between the 24 lowest natural frequencies calculated by 1st, 2nd, 3nd, 4th order approximations and by finite elements method (FEM). Error is calculated with respect to FEM results.

The results of the comparison between the FEM and our approximated solution are outlined in Table 1. It is evident that higher order approximations enhance the accuracy of the solution, as expected. For each mode/row in the table, higher orders leads to increased accuracy.

The characteristic polynomial associated with O(4) is of degree $n^{dof} = 105$. Several roots of this polynomial can be derived analytically, while others need to be found numerically. One of the frequencies which can be obtained analytically is:

$$\omega^2_{fundamental} = \frac{E}{\rho_0 R^2} \frac{45-\sqrt{1605}}{4+4\nu} \tag{47}$$

Importantly, we have found that this frequency is the lowest among all O(4) frequencies in the engineering range of $\nu > 0$. In addition, we have compared between (47) and the fundamental (lowest) frequency obtained by FEM. The maximum difference between the two did not exceed 0.03%. This result is of extreme importance. We have provided, for the first time, a closed form expression for the *fundamental frequency* of a free vibrating elastic 3D cylinder. This expression provides high accuracy for the entire range of practical materials having positive Poisson's ratio.

[43]     E. Hanukah and S. Givli, "Free vibration of an isotropic elastic parallelepiped - a closed form study ".
[44]     P. Wriggers, *Nonlinear finite element methods*: Springer, 2008.
[45]     O. C. Zienkiewicz and R. L. Taylor, *The finite element method for solid and structural mechanics*: Butterworth-Heinemann, 2005.
[46]     Z. Yosibash, *Singularities in elliptic boundary value problems and elasticity and their connection with failure initiation* vol. 37: Springer, 2012.
[47]     P. Shiakolas, K. Lawrence, and R. Nambiar, "Closed-form expressions for the linear and quadratic strain tetrahedral finite elements," *Computers & structures,* vol. 50, pp. 743-747, 1994.
[48]     P. Shiakolas, R. Nambiar, K. Lawrence, and W. Rogers, "Closed-form stiffness matrices for the linear strain and quadratic strain tetrahedron finite elements," *Computers & structures,* vol. 45, pp. 237-242, 1992.
[49]     S. E. McCaslin, P. S. Shiakolas, B. H. Dennis, and K. L. Lawrence, "Closed-form stiffness matrices for higher order tetrahedral finite elements," *Advances in Engineering Software,* vol. 44, pp. 75-79, 2012.
[50]     M. A. Moetakef, K. L. Lawrence, S. P. Joshi, and P. S. Shiakolas, "Closed-form expressions for higher order electroelastic tetrahedral elements," *AIAA Journal,* vol. 33, pp. 136-142, 1995.
[51]     L. Meirovitch and R. Parker, "Fundamentals of vibrations," *Applied Mechanics Reviews,* vol. 54, p. 100, 2001.